# A high-performance deep reservoir computing experimentally demonstrated with ion-gating reservoirs

Daiki Nishioka[1,2], Takashi Tsuchiya[1]*, Masataka Imura[3], Yasuo Koide[4], Tohru Higuchi[2], and Kazuya Terabe[1]

[1]Research Center for Materials Nanoarchitectonics (MANA), National Institute for Materials Science (NIMS), 1-1 Namiki, Tsukuba, Ibaraki, 305-0044, Japan.

[2]Department of Applied Physics, Faculty of Science, Tokyo University of Science, Katsushika, Tokyo 125-8585, Japan

[3]Research Center for Functional Materials, NIMS, 1-1 Namiki, Tsukuba, Ibaraki, 305-0044, Japan.

[4]Research Network and Facility Services Division, NIMS, 1-2-1 Sengen, Tsukuba, Ibaraki, 305-0047, Japan.

*Email: TSUCHIYA.Takashi@nims.go.jp

## Abstract

While physical reservoir computing (PRC) is a promising way to achieve low power consumption neuromorphic computing, its computational performance is still insufficient at a practical level. One promising approach to improving PRC performance is deep reservoir computing (deep-RC), in which the component reservoirs are multi-layered. However, all of the deep-RC schemes reported so far have been effective only for simulation reservoirs and limited PRCs, and there have been no reports of nanodevice implementations. Here, as the first nanodevice implementation of Deep-RC, we report a demonstration of deep physical reservoir computing using an ion gating reservoir (IGR), which is a small and high-performance physical reservoir. While previously reported Deep-RC scheme did not improve the performance of IGR, our Deep-IGR achieved a normalized mean squared error of 0.0092 on a second-order nonlinear autoregressive moving average task, with is the best performance of any physical reservoir so far reported. More importantly, the device outperformed full simulation reservoir computing. The dramatic performance improvement of the IGR with our deep-RC architecture paves the way for high-performance, large-scale, physical neural network devices.

## Introduction

Physical reservoir computing (PRC), which directly utilizes the nonlinear dynamics inherent in physical systems for information processing, has attracted attention in recent years because it can significantly reduce the computational resources required for information processing[1,2]. Nonlinearity, short-term memory, and high dimensionality are required for reservoirs that map input data nonlinearly into a high-dimensional feature space[3,4], and physical devices with these features are promising for PRC[1]. Many types of physical reservoirs have been reported, including memristors, optical devices, spintronics devices, soft bodies, nanowire networks, and ion-gating reservoirs. Further, information processing, including image recognition, spoken digit recognition, and time series prediction tasks, have been demonstrated using such physical reservoir devices[1,2,5–27]. However, since the computational performance of PRCs is still insufficient at a practical level, an approach is required in order to significantly improve performance. One way to achieve this is deep reservoir computing (deep-RC), in which reservoirs are multilayered. This method is expected to be a promising approach, just as neural networks (NN) have been shown to achieve high expressive power and performance by utilizing deep layering. In full-simulation reservoir computing (RC), multilayering of the reservoir parts has been considered, and it has been reported that layering small reservoirs can improve performance in comparison to single-layered reservoirs with the same total reservoir size[28–33]. On the other hand, few attempts at multilayering in physical reservoirs or physical NN have been reported, and those that have been reported are limited to methods that either do not train the connection weights between reservoirs (i.e., the network is not highly flexible)[34,35] or train the connection weights between reservoirs (or layers of physical NN) using backpropagation algorithms that require complex calculations that rely on external circuits and have large computational costs[33,36]. It is particularly notable that there are no reports of deep-RC with nanodevices that are advantageous for integration to realize practical AI devices, and thus, it is not clear that multilayering is effective in improving the performance of physical reservoirs.

Here, as the first implementation of nanodevice-based Deep-RC, we describe a demonstration of deep physical reservoir computing using an ion-gating reservoir (IGR), which is a compact and high-performance physical reservoir. The IGR is a nanodevice with a transistor structure consisting of a hydrogen-terminated diamond channel and a $Li^+$ electrolyte (Li-Si-Zr-O),[23,37] and the ion-electron coupled dynamics based on the EDL effect in the nanoregion near the $Li^+$ electrolyte/diamond interface is used as nonlinear dynamics for the reservoir computing. First, we verified the computational performance of the deep-RC scheme reported for simulation-RC using IGR[28-30]. In this network, the connection weights between reservoir layers are trained based on a simple linear regression algorithm, which provides a higher network flexibility compared to the scheme in which the connection weights between reservoirs are not trained[34,35], and does not require a back-propagation algorithm. However, the conventional scheme with a simple series structure network reported for the simulation-RC does not improve the computational performance of the IGR. This was found to be due to the inherent characteristic of general physical reservoirs, which are sensitive to input conditions. In addition, this scheme does not provide any improvement in network size limitation, which is one of the main reasons why the observed significantly inhibited performance of conventional PRCs. Whereas in simulation-RC it is easy to increase the reservoir size to the desired performance in exchange for computational cost, in PRCs the number of reservoir states (current and voltage response, mechanical vibration, optical response, etc.) obtained from the physical device is limited by the means of access to the physical system, such as measurement probes. Thus, it is difficult to increase the network size of such PRCs.

On the other hand, a deep-RC scheme with parallel structure overcome such limitations of physical reservoirs in principle, and succeeded in greatly increasing the number of reservoir states by utilizing the reservoir outputs of the previous layer as well as the final outputs among the layered device outputs. As a result, the performance of the IGR applied with the subject deep-RC scheme (deep-IGR) was significantly improved, with a 41% reduction in error compared to single-layer IGR in a second-order nonlinear autoregressive moving average (NARMA2) task. However, the model also showed an increase of unnecessary reservoir states leading to overlearning. Therefore, a modified model was developed to improve this model by evaluating the high dimensionality of the reservoir state in terms of the correlation coefficient between reservoir states, the number of layers was increased by excluding featureless reservoir states (which do not contribute to high dimensionality), which resulted in a 53% reduction in error for the NARMA2 task compared to single-layer IGR, with a normalized mean squared error (NMSE) of 0.0092. This is the best performance of any physical reservoir reported to date [22–25,38–40], outperforming a full-simulation RC [38].

## Results

### Ion-gating reservoirs using electric double-layer transistors

To demonstrate the applicability of the deep-RC architecture, we employed an IGR as the physical reservoir part, implemented with an electric double-layer (EDL) transistor composed of a $Li^+$ electrolyte (Li-Si-Zr-O) and hydrogen-terminated diamond, as shown in Fig. 1**a**. The IGR transistor has 9 channels of different lengths $L_{ch}$ (=5, 10, 25, 35, 50, 100, 250, 350, 500 μm), and the 9 drain current responses can be nonlinearly transformed by the EDL mechanism to the input gate voltage signal. Figure 1**b** shows the drain current ($I_D$)-gate voltage ($V_G$) curves obtained from channels with lengths of $L_{ch}$=5 μm and $L_{ch}$=500 μm (upper panel) and a gate current ($I_G$) – $V_G$ curve (lower panel). When a positive gate voltage is applied, $Li^+$ in the electrolyte accumulate on the diamond surface, forming an EDL at the electrolyte/diamond interface. Then, electrons are doped into the hydrogen-terminated diamond channel, which is a hole conductor, and the drain current is modulated nonlinearly[37]. As different channel resistances exhibit different relaxation times for different channel lengths, if the gate input is a pulse signal, it will exhibit different drain current responses depending on the channel length, as shown in Fig. 1**c**, providing a higher dimension as a physical node[23]. In addition to these drain currents, the gate currents, obtained from the input gate terminals, show a spiked response, as shown in the bottom panel of Fig. 1**c**, which provides additional diversity to the IGR[24,26].

In redox-based IGR, the use of gate currents as reservoir states has been reported to significantly improve computational performance[24]. In addition to the 10 physical nodes obtained by adding the gate current response to the 9 drain current responses, 10 nodes per pulse response were obtained as virtual nodes, as shown in the right-hand panel of Fig. 1**c**. Thus, the number of reservoir states $X_i$ obtained from the IGR (i.e., nodes) is 100. Said reservoir states $X_i$ were normalized from 0 to 1 and subsequently used for the reservoir part in the schematic of reservoir computing shown in Fig. 1**d**. The reservoir output $y(k)$ at discrete time step $k$ is obtained by the linear combination of reservoir state $X_i(k)$ and readout weight $W_i$ as follows;

$$y(k) = \sum_{i=1}^{N} W_i X_i(k) + b \qquad (1)$$

,where $b$ and $N$ are the bias and reservoir size, respectively. Figure 1**e** is a schematic of the deep-IGR which is a physically implemented deep reservoir computing with IGRs. In the first layer, as in conventional IGRs [23], the voltage-transformed input signal was input to the gate of the IGR, and the reservoir output was calculated by a linear sum of the weights and reservoir states (Eq. 1) obtained by acquiring virtual nodes from the obtained current responses, as shown in Fig. 1**c**. The weights were trained by linear regression so that the target and reservoir outputs matched. In deep-IGR, the reservoir output of the first layer is converted to a voltage pulse and input to the gate of the second layer IGR, and the reservoir output is obtained via weights using the obtained current response as in the first layer. The input that reproduced the target waveform to some degree in the first layer is again nonlinearly transformed by IGR into a higher dimensional feature space for learning, which allows the representation of target features that could not be represented in the first layer. The number of said layers can be increased by inputting the reservoir outputs of the previous layer, in the same way as in the second layer, for the third and subsequent layers. The performance of deep-IGR was evaluated by the error between the target and the reservoir output. obtained by the input and forward propagation of a dataset different from the training dataset, with the weights of all layers fixed. The details of deep-IGR are discussed in Figure 3, and later in this document.

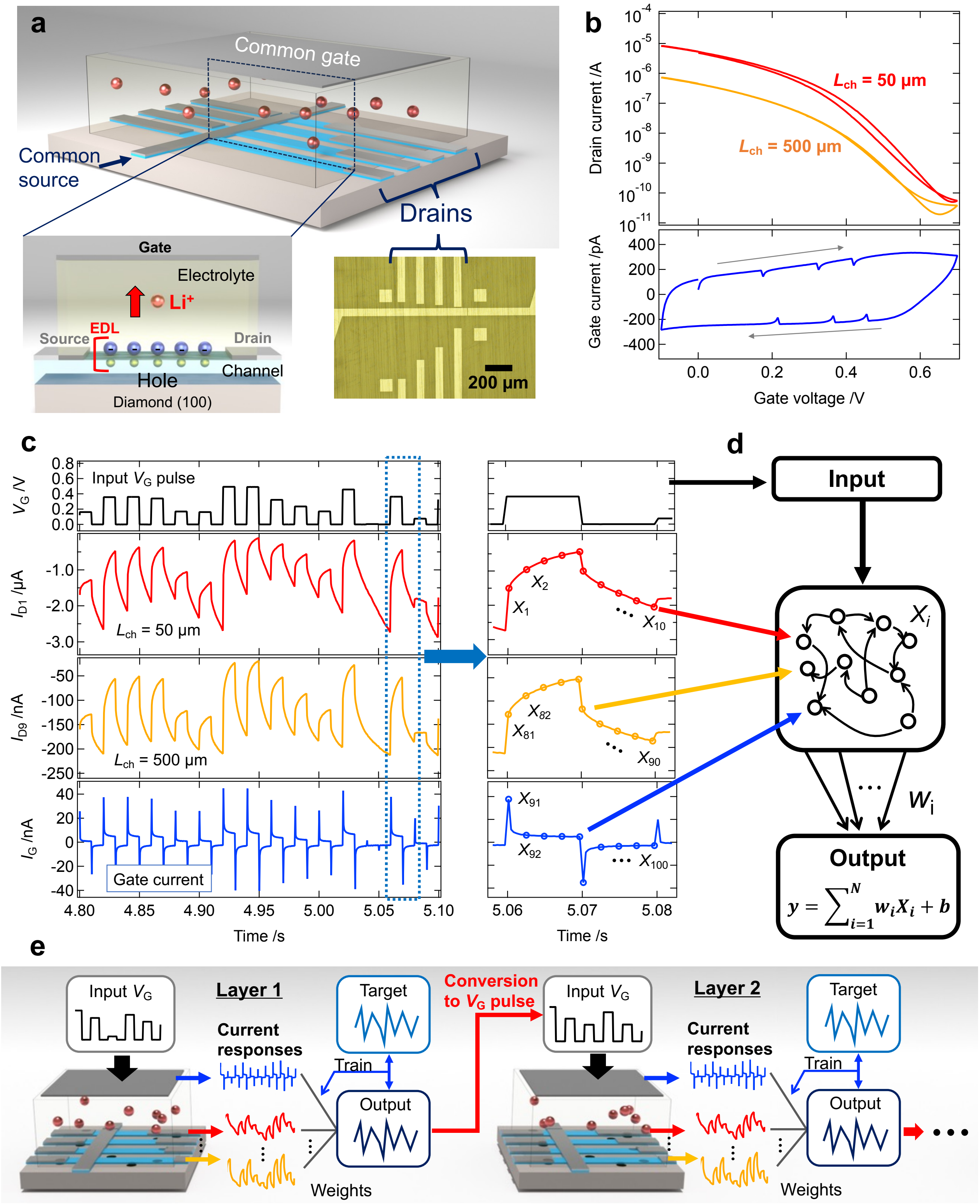

**Figure 1. a** Schematic of the IGR transistor. The inset is an optical microscope image of a diamond with source and drain electrodes. **b** $I_D$-$V_G$ curves (upper panel) for channels with lengths of 50 μm and 500 μm and $I_G$-$V_G$ curve (lower panel). **c** $I_D$ responses ($L_{ch}$=50μm, 500μm) and $I_G$ response to pulsed $V_G$ input. The right-hand panel shows how to obtain virtual nodes from such current responses. **d** Schematic of reservoir computing. **e** Schematic of the deep reservoir computing implemented by IGRs.

Before evaluating the performance of deep-IGR, we performed a NARMA2 task that predicts the NARMA2 model shown in Eq. 2, in order to evaluate the computational performance of the single-layer IGR.

$$y_t(k+1) = 0.4y_t(k) + 0.4y_t(k)y_t(k-1) + 0.6u^3(k) + 0.1 \quad (2)$$

where $y_t(k)$ and $u(k)$ are the model outputs and random inputs, ranging from 0 to 0.5, respectively. The NARMA2 task, which is widely used as a benchmark task for physical reservoirs, requires reservoirs to exhibit second-order nonlinearities and short-term memory [22–25,38–40]. Figure 2**a** shows a schematic of the NARMA2 task performed by IGR. The random input signal $u(k)$ was converted into voltage pulse streams with an intensity of 0 V to 0.5 V, a pulse period of $T$, and a duty cycle of $D$, then input to the gate terminal of the IGR transistor. The responses of the drain current to the gate voltage pulse streams were measured at a constant drain voltage of -0.5 V. As shown in Fig. 1**c**, 100 reservoir states $X_i$ were obtained from 10 current responses and virtual nodes ($i$=1, 2, …, 100). Furthermore, an additional 100 reservoir states $X_i$ ($i$=101, 102, …, 200) were obtained by applying an intensity-reversed input $u_{\mathrm{inv}}(k)$ [= 0.5 - $u(k)$] to the IGR (inversion pulse method), resulting in a total of 200 reservoir states, the combination of which was utilized to obtain the reservoir output as shown in Eq. 1. Details on the inversion pulse method are given elsewhere [23]. In the training phase, the readout weights were trained by linear regression, in order to match the reservoir output $y(k)$ and the model output $y_t(k)$. Details of the training algorithm are given in the Method section herein. In the test phase, performance was evaluated by NMSE (Eq. 3) of the reservoir output (prediction) $y(k)$ to the model output $y_t(k)$ for a different input $u(k)$ than in the training phase.

$$\mathrm{NMSE} = \frac{1}{M}\frac{\sum_{k=1}^{M}[y_t(k) - y(k)]^2}{\sigma^2[y_t(k)]} \quad (3)$$

, where $M$ and $\sigma^2$ are the data length ($M$=1600 for the training phase and $M$=700 for the test phase) and variance, respectively. Figures 2**b** and **c** show the pulse period $T$ and duty cycle $D$ dependence of NMSEs in the training and test phases, respectively. The best results, for both training and test phases, were obtained at $T$=70 ms and $D$=70 %, where the NMSE was 0.0157 in the training phase and 0.0194 in the test phase. Fig. 2**d** shows the $D$ dependence of NMSE at $T$=70 ms, and Fig. 2**e** shows the $T$ dependence of NMSE at $D$=70 %. Both of these are minima at the optimal conditions (indicated by *), indicating that the search for optimal conditions in single-layer IGR was successfully performed. In other words, this is the limit of the computational performance that can be achieved by adjusting the pulse period and duty cycle. Further performance improvements require consideration of voltage conditions ($V_G$ and $V_D$) and preprocessing of the input signal (feature extraction, masking, etc.). Tuning these so-called ‘hyperparameters’ is difficult for physical reservoirs that require actual measurements, and the huge variety of parameters (voltage, time, temperature, number of masks, etc.) and their combinations make it extremely difficult to maximize the potential computational performance of the physical system. In all of the deep-IGR experiments discussed below, the voltage pulse conditions were fixed at $T$=70 ms and $D$=70%. In this case, the optimal conditions match between training and testing, but if such is not the case, the optimal conditions in the training data should be adopted as the input conditions for the second and subsequent layers in order to avoid incorrect optimization by the testing data.

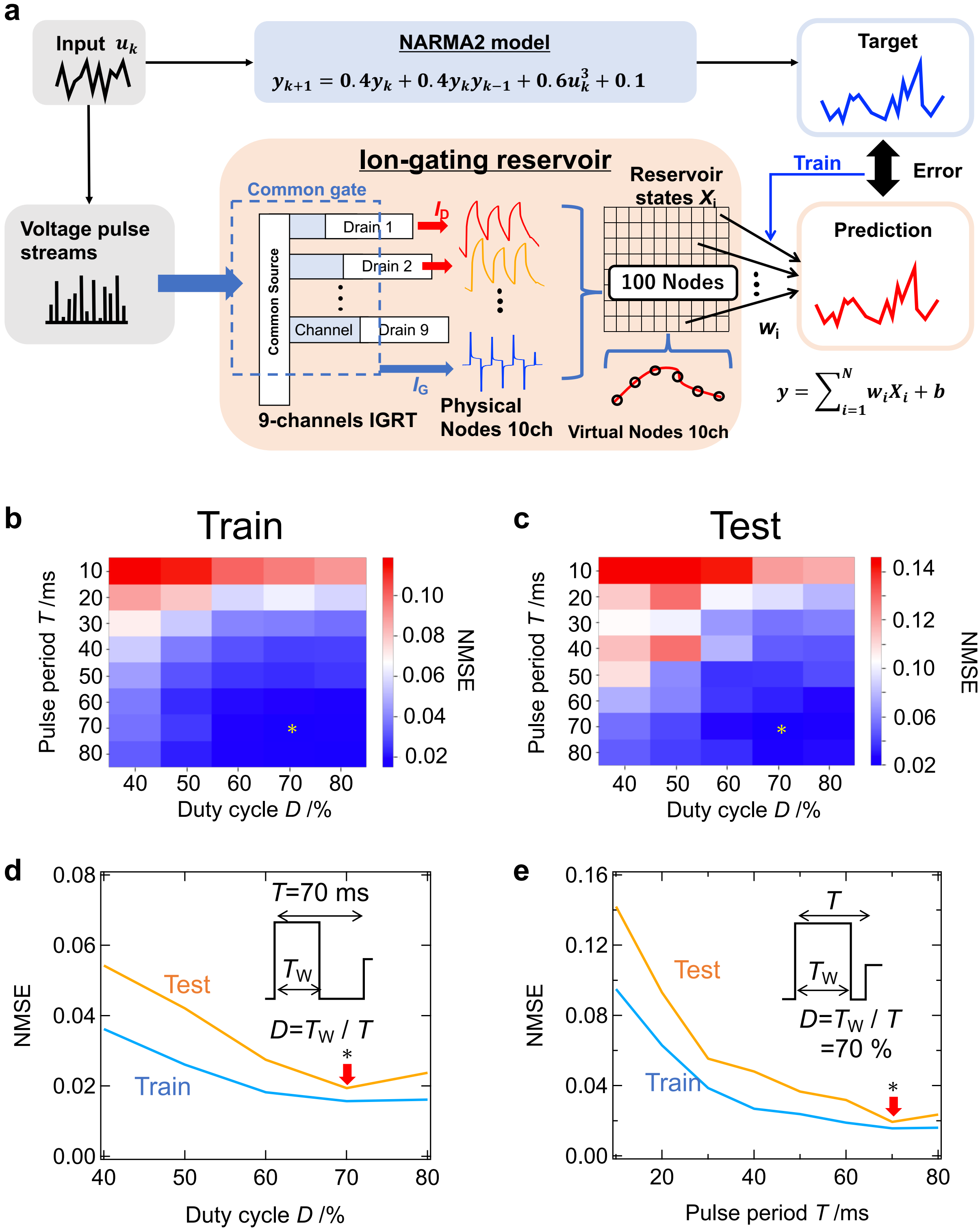

**Figure 2. a** Schematic of the NARMA2 task performed by IGR, showing the pulse period *T* and duty cycle *D* dependence of NMSEs for training (b) and test (c) phases. **d** D-dependence of NMSEs at *T*=70 ms. **e** *T*-dependence of NMSEs at *D*=70 %. Optimal conditions are indicated by *.

## Performance evaluation of deep ion-gating reservoir with NARMA2 task

We experimentally demonstrated the deep-IGR shown in Figs 3**a**,**b** as a framework that overcomes the limitations discussed above, and easily maximizes the computational performance of the physical system. Although the deep layered network structure shown in Fig. 3**a** (Network 1) has been reported to improve performance in a full simulation reservoir [28–30], there are no reports of its application to a physical reservoir. In the first layer of Network 1, the readout weight $W^{(1)}$ is trained with the same procedure as with the single-layer IGR shown in Fig. 2**a,** so as to obtain the first-layer reservoir output $y^{(1)}(k)$. In the $L$-th layer($L \geqq 2$), the voltage-transformed reservoir output $y^{(L-1)}(k)$ from the previous layer is input to the IGR instead of random input $u(k)$. The reservoir output of the $L$-th layer is then calculated by the linear combination of the reservoir state $X^{(L)}(k)$ obtained from the IGR and the readout weight $W^{(L)}$ trained by linear regression. In the test phase, the signal was propagated forward with the weights of all layers fixed, and the computational performance was evaluated by the error between the reservoir output $y^{(L)}(k)$ and the model output $y_t(k)$ (Eq. 2) at the final layer $L$. The black and gray plots in Fig. 3**c** show the dependence of the NMSE on the number of layers in the test phase and the training phase, respectively, for the NARMA2 task performed on Network 1 to the training data. In this task, the reservoir is used to predict the NARMA2 model output $y_t(k)$ from input $u(k)$. However, in the structure of Network 1 (deep layered), the nature of the problem changes after the second layer, and the task switches to predicting the NARMA2 model output using the reservoir prediction of the previous layer as input (task switching). In a physical reservoir that uses the transient response of a physical system in real time, the properties of the reservoir state (nonlinearity, memory capacity, and diversity) change significantly depending on the operating conditions (in this case, the input $V_G$ pulse stream conditions), so that the accuracy in a given task changes due to sensitivity to the device operating conditions (i.e., as discussed in Fig. 2, the performance varies greatly with the operating conditions of the IGR.). Therefore, in this case, where the nature of the task has changed, different operating conditions for the IGR need to be explored because different characteristics are required for the reservoir. However, the method of searching for optimal operating conditions for each additional layer is not realistically achievable.

Therefore, in order to overcome such drawback and improve performance, without the need to search for optimal operating conditions, we considered Network 2 (Fig. 3**b**). This network is a modified model of Network1 (deep layered) that utilizes all the reservoir states obtained from the reservoirs of the previous layers $[X^{(1)}(k), X^{(2)}(k),\ldots, X^{(L)}(k)]$ so as to obtain the reservoir output of a given $L$-th layer (deep and parallel structure). In this scheme, higher dimensionality (the number of reservoir states used to obtain output) increases with the number of layers, so that the expressiveness of the reservoir also increases. Another major difference from Network 1 (deep layered) is that task switching does not occur as discussed in Network 1. Since the reservoir state $X^{(1)}(k)$ in the first layer is always used in the computation, regardless of the number of layers, the reservoir states in the second and subsequent layers can be interpreted as additional features that improve the accuracy of predicting the NARMA2 model from the reservoir states in the first layer. The blue and light blue plots show the dependence of the NMSE on the number of layers in the test phase and the training phase, respectively, for the NARMA2 task performed on Network 2 (deep and parallel layered)[28]. The NMSE decreases significantly as the number of layers increases, and the error decreases by 41% at the third layer compared to the single layer. Figure 3**d** shows the relationship between NMSEs and the number of reservoir states used to generate output in each layer ($N_{\mathrm{OUT}}$) for the two networks. Network 2 (deep and parallel layered) clearly shows a reduction in errors compared to Network 1(deep layered). This is possibly due to the fact that the number of nodes was effectively increased, while such increase is generally difficult to achieve with physical reservoirs, as mentioned above. However, in Network 2 (deep and parallel layered), the error increased slightly at the fourth layer, which increase effectively halted the performance improvement.

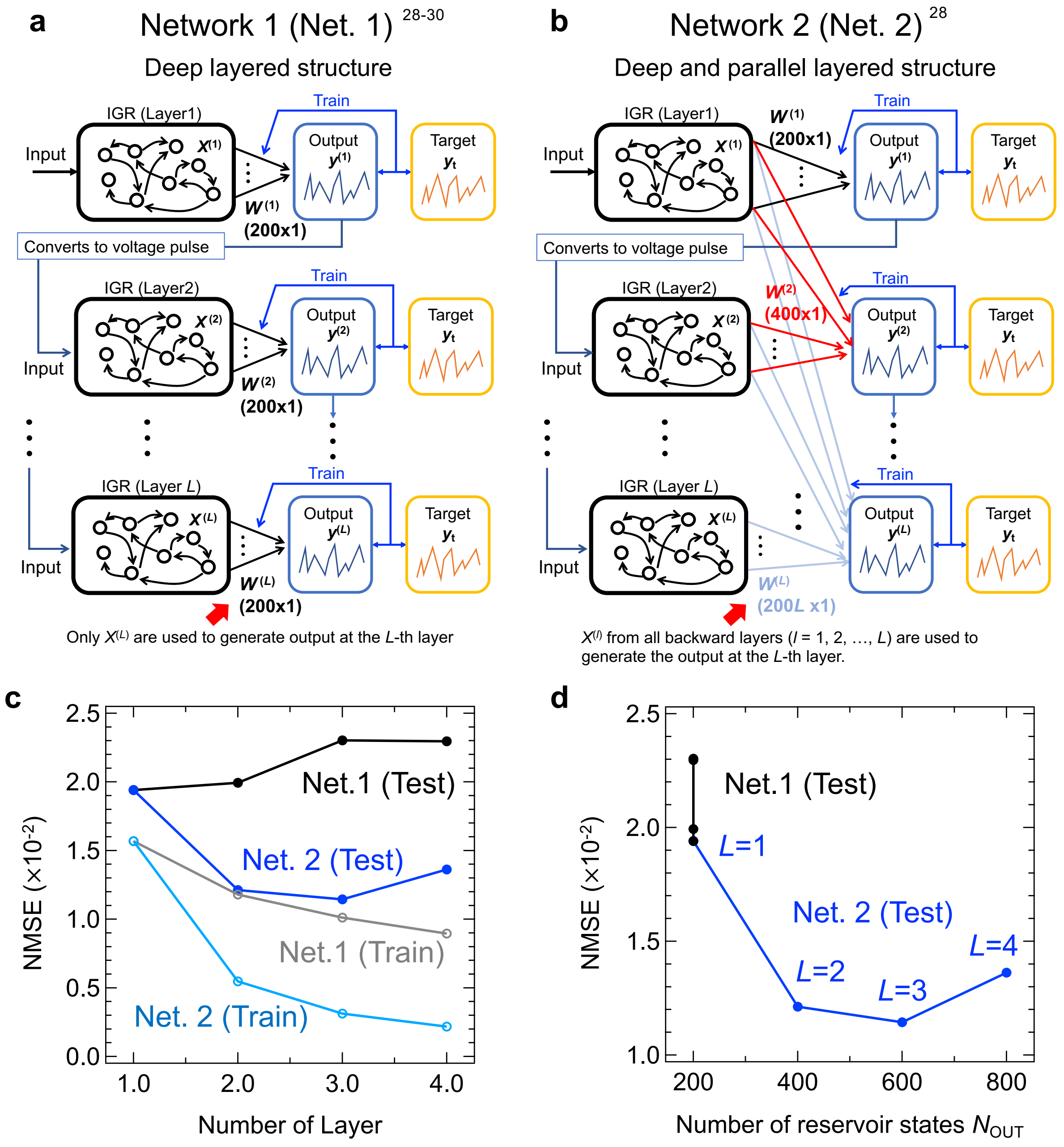

**Figure 3.** Schematic diagram of deep-IGR for (a) Network 1 [28-30] and (b) Network 2 [28]. **c** NMSEs of the NARMA2 task vs. the number of layers of deep-IGR. The black and blue plots show the results for Network 1 and Network 2, respectively. **d** NMSEs of the NARMA2 task vs the number of reservoir states used to generate output in each layer.

**Node Selection in deep-IGR**

The decrease in performance with respect to the increase in the number of nodes, which is shown in Fig. 3d, is thought to originate from the effect of overfitting due to the increase in the number of unnecessary nodes[41]. To verify this hypothesis, we analyzed the high dimensionality in the correlation coefficient $r_{AB}$ between node A and node B shown in Eq. 4 for the reservoir state used at the output of the fourth layer of this network 2[24].

$$r_{AB} = \frac{\sum_{k=1}^{M}(X_A(k) - \bar{X}_A)(X_B(k) - \bar{X}_B)}{\sqrt{\sum_{k=1}^{M}(X_A(k) - \bar{X}_A)^2 \sum_{k=1}^{M}(X_B(k) - \bar{X}_B)^2}} \tag{4}$$

, where $X_A(k)$ is the reservoir state of node A and $\bar{X}_A$ is the average value of $X_A(k)$. Figure 4**a** shows the correlation coefficients $|r_{AB}|$ between all reservoir states $X_i^{(L)}$ ($i$=1~200, $L$=1~4) used to generate output in the fourth layer of Network 2 (i.e., 800 nodes in total). In the first layer, a random wave $u(k)$ was input to the IGR, whereas in layer $L$ ($\geqq$2), the output $y^{(L-1)}(k)$ of the previous layer was input to the IGR. Therefore, because these correlation coefficients are small, due to the difference in inputs, the $|r_{AB}|$ in the first layer is very different from that in the rest of the layers, as shown in Fig. 4**a**. The inset of Fig. 4**a** shows an expanded part of the reservoir state in the first layer. For example, the correlation coefficient $|r|$ between $X_{i=5}^{(L=1)}$ and $X_{i=3}^{(L=1)}$ is close to 1, and they exhibit similar behavior, as shown in Fig. 4**b**. They correspond to different virtual nodes taken from the same drain current ($L_{ch}$=5 μm). On the other hand, $X_{i=5}^{(L=1)}$ and $X_{i=11}^{(L=1)}$ have a relatively low correlation coefficient $|r|$ of 0.48 and exhibit different behavior, as shown in Fig. 4**c**. These nodes correspond to different virtual nodes taken from different drain currents ($L_{ch}$=5 μm and $L_{ch}$=10 μm).

To identify which of the 800 nodes shown in Fig. 4**a** has the lower correlation coefficient (i.e., contributes to the higher dimensionality), the average value of the correlation coefficient for node $j$ (=1, …, 800) was calculated as shown in Eq. 5;

$$|\bar{r_j}| = \frac{\sum_{i\neq j}^{N_{all}}|r_{ji}|}{N_{all} - 1} \tag{5}$$

, where $N_{all}$ is the number of all nodes (in this case, $N_{all}$ = 800). The $|\bar{r_j}|$ plots, rearranging the node numbers $j$ in order from the lowest $|\bar{r_j}|$, are shown in Fig. 4**d**. The lowest $|\bar{r_1}|$ is 0.3, whereas $|\bar{r_j}|$ increases with increasing $j$, indicating that $|\bar{r_j}|$ saturates at about $j$=400. This suggests that about 400 of the total 800 node reservoir states are nodes that do not contribute to high dimensionality (i.e., the nodes are less effective for performing the given task). To evaluate the effect of node correlation coefficients and high dimensionality on computational performance, the NARMA2 task was performed by increasing the reservoir size $N'$ in the order of lower $|\bar{r_j}|$ (for example, when $N'$=100, $X_j$($j$=1~100) was used for the calculation). Figure 4**e** shows the relationship between $N'$ and NMSE; in the region where $N'$ is approximately 400 or less, $N'$ increases while NMSE decreases for both training and testing. On the other hand, in the region where $N'$ is above approximately 400, $N'$ increases and the training error continues to decrease, while the test error increases. This indicates the effect of over-fitting with increasing reservoir size, in which reservoir states in regions of saturated diversity have a negative impact on the computation. Figure 4**f** shows the results obtained with size reduction (Fig. 4**e**) and the relationship between the NMSEs of the NARMA2 task (test phase) and $N_{\mathrm{OUT}}$ for Network 2. The red plots show the results of the computation when gradually excluding nodes, in order of largest $|\overline{r_j}|$, from the 800 nodes which were used to generate output in the fourth layer of Network 2. It was found that the NMSE decreased despite any reduction in the number of reservoir states used to generate the output. Reducing to $N'$=600 (i.e., excluding from the calculation the 200 nodes with large $|\overline{r_j}|$), the improvement in computational performance was only about the same as in the third layer without size reduction, but at $N'$=400 (i.e., excluding from the calculation the 400 nodes with large $|\overline{r_j}|$), it outperformed Network 2 without size reduction. The performance continued to improve as the number of nodes was reduced to about $N'$=300, but when $N'$ was further reduced, the performance rapidly degraded due to loss of the nodes necessary for the computation being performed. These results indicate that the adverse effects of multilayering, such as unnecessary node growth and overfitting, were successfully suppressed without sacrificing the advantages of improved computational performance due to multilayering.

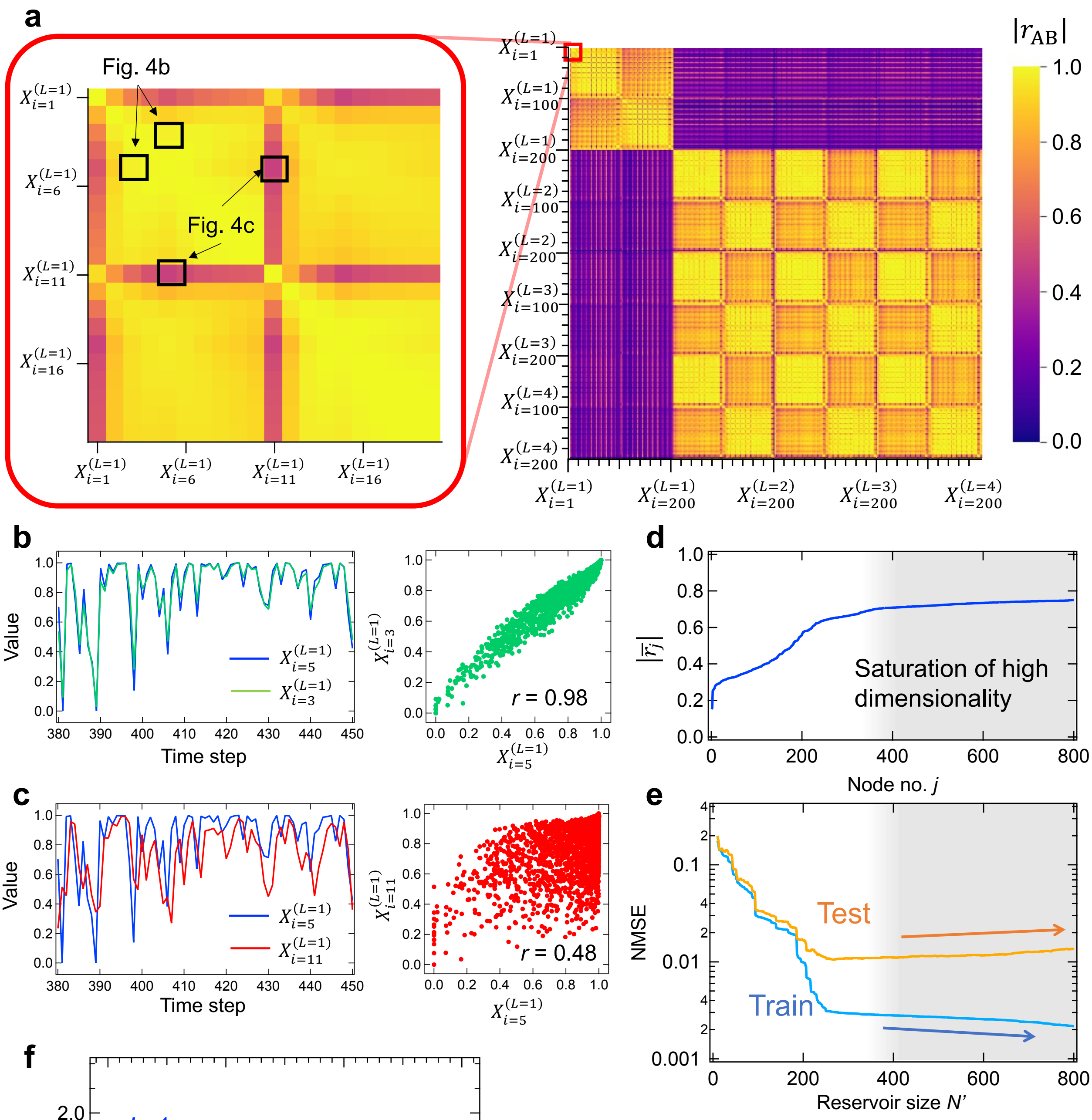

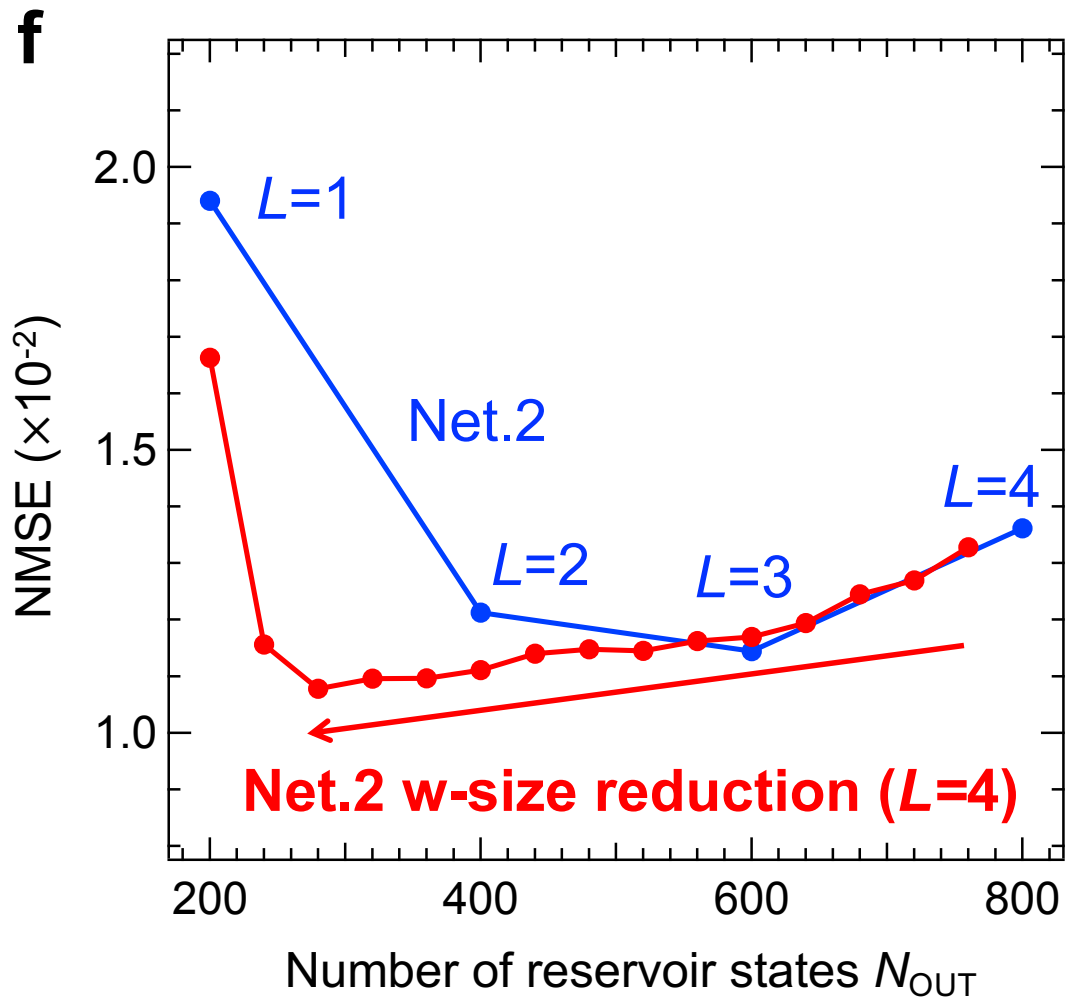

**Figure 4. a** Heatmap of the correlation coefficients of the reservoir states used in the output for the 4th layer of Network 2. An example of (**b**) a strongly correlated or (**c**) relatively weakly correlated reservoir state and a scatterplot. **d** The $|\overline{r_j}|$ plots, rearranging the node numbers $j$ in order from the lowest $|\overline{r_j}|$. **e** The relationship between the number of selected nodes $N'$ and the NMSE of the NARMA2 task. **f** $N_{OUT}$ vs NMSEs of NARMA2

**Deep-IGR architecture, which utilizes node selection at each layer**

The node selection (reduction) by $|\overline{r_j}|$ was adapted to only one layer in the configuration shown in Fig. 4. Here, in order to maximize computational performance, we consider Network 3, in which node selection is adapted to all layers from the second layer onward by modifying Network 2 (deep and parallel layered). The number of selected reservoir states $N'$ used for the output of each layer was set to 400. Therefore, the first (single) and second layers, where the number of reservoir states $N_{\mathrm{OUT}}$ is 200 and 400, respectively, do not perform node selection and are therefore identical to Network 2, as shown in Fig. 3**b**. The calculation method used for the third and subsequent layers of Network 3 is explained below. The output in the third layer of Network 2 used 600 reservoir states of $X^{(L=1)}$, $X^{(L=2)}$ and $X^{(L=3)}$ [here, these reservoir states are rewritten as $X_j$ ($j$=1, 2,…, 600)]. In Network 3, of these 600 reservoir states, $N'$ nodes (here 400 nodes) selected in order of smallest $|\overline{r_j}|$ are used to generate the output of the third layer $y^{(L=3)}$. Furthermore, by voltage converting $y^{(L=3)}$ and inputting it to IGR again, $X^{(L=4)}$ is obtained, and together with the reservoir states $X^{(L=1)}$, $X^{(L=2)}$ and $X^{(L=3)}$ obtained in the previous layers, there are a total of 800 reservoir states [here, these reservoir states are rewritten as $X_j$ ($j$=1, 2,…, 800)]. As above, $N'$ (=400) of these 800 nodes with small $|\overline{r_j}|$ were selected $X'^{(L=4)}$ and used to generate output in the fourth layer $y^{(L=4)}$. Figures 5**a** and **b** show schematic diagrams of how the reservoir outputs $y^{(L)}$ of Network 2 and Network 3 in layer $L$ are computed, respectively. In Network 2, shown in Fig. 5**a**, the number of reservoir states used for output increases by 200 as the number of layers increases, because all reservoir states $X^{(L=1)}$, $X^{(L=2)}$, …, $X^{(L)}$ in layers 1 to $L$ are used to generate the reservoir output in layer $L$. On the other hand, in Network 3, shown in Fig. 5**b**, $X'^{(L)}$ selected by $N'$ nodes in order of smallest $|\overline{r_j}|$ of the reservoir states $X^{(L=1)}$, $X^{(L=2)}$, …, $X^{(L)}$ in layers 1 to $L$ is used for the output in layer $L$, as described above. Therefore, the number of reservoir states used to generate output is fixed at $N'$, regardless of the number of layers $L$ (>1).

Figure 5**c** shows a plot of NMSE vs. the number of layers in the NARMA2 task for Network 3 (deep and parallel layered with node selection), with the number of selected nodes $N'$=400. The NMSE continued to decrease as the number of layers increased, for both training and testing errors, with the fourth layer achieving better computational performance than Network 2 (deep and parallel layered), shown by the dotted blue line, with an NMSE of 0.00326 in the training phase and 0.00921 in the testing phase. These values are 79% and 53% lower in the training and testing phases, respectively, compared to a single-IGR. Figure 5**d** shows the relationship between the number of nodes used for output $N_{\mathrm{OUT}}$ and NMSE (test phase). Network 3 showed improved computational performance over the conventional Network 2, despite $N_{\mathrm{OUT}}$ after the second layer being fixed at 400 (=$N'$). This performance improvement is explained by the network being trained by increasing the number of layers while excluding unnecessary nodes that cause overfitting and nodes that do not contribute to diversity, thereby effectively incorporating the features obtained in each layer. The predicted and target waveforms at layers 1 and 4 are shown in Figs. 5**e** and **f**, respectively. Although the predicted waveform in the first layer captured the trend of the target waveform, there was a divergence between the predicted waveform and the target waveform in some areas. On the other hand, the predicted

waveforms at the four layers, shown in Fig. 5**f**, are in almost perfect agreement. This shows that the deep-IGR based on Network3 is able to successfully solve the NARMA2 model shown in Eq. 2, and that this architecture significantly improved the computing performance of the IGR.

Figure 6**a** shows a comparison of the computational performance evaluated in the NARMA2 task of this study with other physical reservoirs reported so far [22–25,38–40]. Our deep-IGR (Network 3, $L$=4) achieved NMSE=0.00921 in the test phase for the NARMA2 task, and achieved the highest computational performance of any physical reservoir reported to date [22–25,38–40], even outperforming the NMSE=0.016 of fully simulated reservoir computing [Standard Echo State Network (ESN)][38]. The deep-RC architecture proposed in this study has very few hyperparameters, and can easily enhance the computational performance of physical reservoirs. Usually, it is extremely difficult to maximize the information processing capability of a physical system in physical reservoir computing. There are a wide variety of parameters to be considered if the best computational performance is to be obtained from a physical reservoir; these include the delay time of the feedback loop, the mask matrix used for input signal preprocessing (mask format, number of masks, mask length, etc.), in addition to the intensity and frequency of the signal input to the physical system. Such complexity makes it almost impossible to experimentally search for the best combination of driving conditions for a physical system. Hence, the deep-layered physical reservoir architecture proposed in this study can dramatically improve the computational performance, and can be realized in a small parameter space that is possible to realistically explore. We experimentally examined three different deep reservoir architectures, which are shown in Figs. 6**b-d**, and found that Network 1 (deep layered, shown in Fig. 6**b**), which was previously reported for simulated reservoirs, does not necessarily contribute to improved performance in physical reservoirs [28–30]. Said scheme is suitable only as a component part of physical reservoirs, whose characteristics are not so sensitive to operating conditions. Further, it does not increase the network size, which is a serious problem faced by physical reservoirs. On the other hand, we have succeeded in increasing the network size and in improving the computational performance of the physical reservoir significantly with Network 3 (deep and parallel layered with node selection), which is a modified version of Network 2 (deep and parallel layered, shown in Fig. 6**c**) that selectively uses useful nodes for information processing, as shown in Fig. 6**d**. Our deep-IGR is the first nanodevice implementation of deep-RC, and the dramatic improvement of the performance of IGR by our architecture provides the possibility of realizing large-scale, brain-like physical devices.

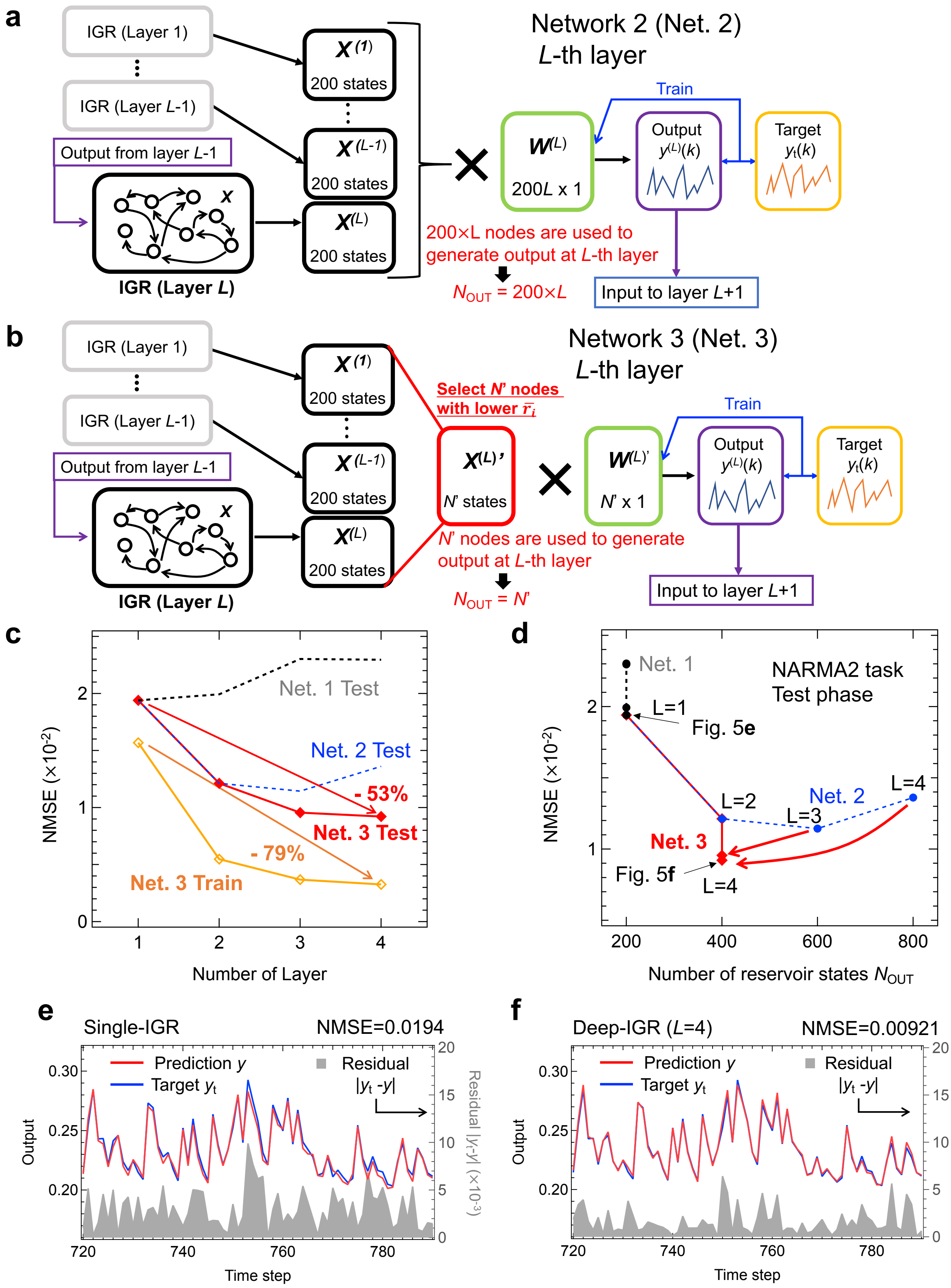

**Figure 5.** Schematic diagram of layer $L$ of a deep-IGR for (**a**) Network 2 and (**b**) Network 3 with node selection. **c** NMSEs of the NARMA2 task vs. the number of layers of deep-IGR. **d** NMSEs of the NARMA2 task vs $N_{\mathrm{OUT}}$. Target waveform and predicted waveform by IGR at (**e**) $L$=1 and (**f**) $L$=4

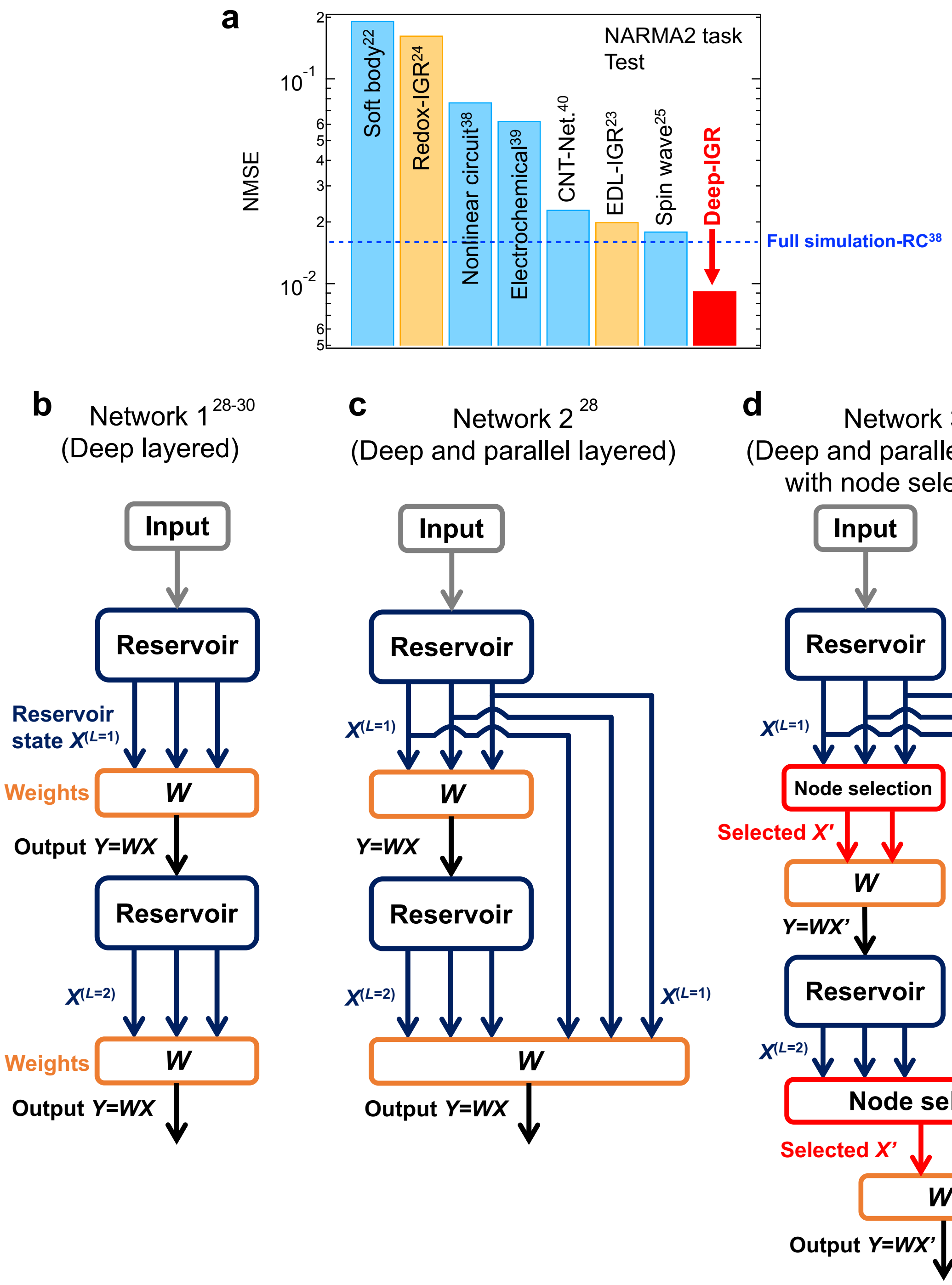

**Figure 6. a** Performance comparison with other physical reservoirs by NARMA2 task[22-25,38-40]. The result for soft body[22] was converted to the NMSE in Eq. 3 for comparison. Schematic diagram of the Deep-RC architecture for PRC investigated in this study for (b) Network 1 [28-30], (c) Network 2 [28] and (d) Network 3.

## Conclusion

In this study, we demonstrated implementation of Deep-RC by nanodevices for the first time, whereas Deep-RC has only been implemented by simulated RCs and limited physical reservoirs. We implemented IGR that uses ion-electron coupling dynamics as a computational resource [23] in the reservoir part, so as to create a multi-layered framework, and found that a simple serial network structure did not improve the performance (Network 1). On the other hand, we determined that a network structure that uses the reservoir state of the previous layer for outputs shows a significant improvement in performance (Network 2). This is because, in addition to succeeding in further increasing the dimension, which is generally difficult in physical systems, the predicted output of the reservoir is again input to the IGR and transferred to the high-dimensional feature space, where it is learned so that the deviation from the target output is reduced. In order to suppress overlearning due to increase of unnecessary nodes, which was confirmed in this network structure, node selection based on correlation coefficients was used in Network 3 to effectively extract nodes that are effective for information processing. As a result, an NMSE of 0.00921 was achieved in the NARMA2 task for deep-IGR. This is the best performance of all the physical reservoirs reported so far, and even outperforms full-simulation RC [22–25,38–40]. The easy and dramatic performance improvement of IGRs with our deep-RC architecture, and opens the way to the implementation of practical, large-scale, brain-based physical devices.

## Method

### Device fabrication and electrical measurements

Hydrogen-terminated diamond homoepitaxial film was deposited, as a channel, on a IIa-type high-pressure high-temperature single crystal diamond substrate (100) (EDP) by the microwave-plasma chemical vapor deposition (MPCVD) method. During the deposition process, 500 and 0.5 standard cubic centimeters per minute of $H_2$ and $CH_4$, respectively, were supplied, and the hydrogen-terminated diamond was grown at a radio frequency of 950 W. The IGR were fabricated with nine different channel lengths (5, 10, 25, 35, 50, 100, 250, 350, and 500 μm), all with a channel width of 50 μm. Pd/Pt electrodes (10 and 35 nm, respectively) were deposited as source and drain electrodes by electron beam evaporation with maskless lithography after oxygen termination of the diamond surfaces (other than channels) by oxygen plasma asher. A 3.5-μm LSZO thin film, used as an electrolyte, was deposited by pulsed laser deposition (PLD) with an ArF excimer laser. A 100-nm Au thin film was deposited by electron beam deposition as a gate electrode.

Electrical measurements of IGR were performed by the source measure unit and pulse measure unit of a semiconductor parameter analyzer (4200A-SCS, Keithley), which measurements were carried out at room temperature inside a vacuum chamber that had been evacuated by a turbo molecular pump. Probers were used to connect the IGR inside the chamber.

**Linear regression algorithm for learning readout networks**

In the NARMA2 task, which was used in this study to evaluate performance, the readout network was trained with linear regression using the algorithm described below. The reservoir output shown in Eq. 1 can also be expressed as;

$$y(k) = \boldsymbol{W} \cdot \boldsymbol{x}(k) \tag{6}$$

where $W = (b, W_1, W_2, \ldots, W_N)$ and $x(k) = [1, X_1(k), X_2(k), \ldots, X_N(k)]^T$ are the weight vector and the reservoir state vector, respectively. The reservoir output $Y$ for all training periods ($k$=1, 2, …, $M$) is described by;

$$Y = WX \tag{7}$$

where $X = (\boldsymbol{x}(1), \boldsymbol{x}(2), \ldots, \boldsymbol{x}(M))$ and $M$ = 1600 are the reservoir state matrix and the data length for the training phase, respectively. The weight matrix that minimizes the squared error is given by;

$$W = Y_t X^\dagger \tag{8}$$

where $X^\dagger = [X^T(XX^T)^{-1}]$ and $Y_t$ are the Moore-Penrose pseudo-inverse matrix and the target matrix, respectively. The learning and the sum-of-product calculation performed in the readout network were done on a personal computer using current data obtained from the IGR. However, by utilizing artificial synaptic devices that reproduce weights by conductance, the sum-of-product calculation performed in the readout can also be calculated in a physical process, which is expected to further improve efficiency[42-51].

**Acknowledgement:**

This work was in part supported by Japan Society for the Promotion of Science (JSPS) KAKENHI Grant Number JP22H04625 (Grant-in-Aid for Scientific Research on Innovative Areas "Interface Ionics"), and JP22KJ2799 (Grant-in-Aid for JSPS Fellows). A part of this work was supported by the Iketani Science and Technology Foundation.

**Author contributions**:

D.N., T.T., and K.T. conceived the idea for the study. D.N. and T.T. designed the experiments. D.N. and T.T. wrote the paper. D.N. carried out the experiments. D.N. prepared the samples. D.N. and T.T. analyzed the data. All authors discussed the results and commented on the manuscript. K.T. directed the projects.